\documentclass[prl,tightenlines,twocolumn,showpacs,preprintnumbers,nobibnotes,aps,superscriptaddress,10pt]{revtex4-1}

\usepackage{amsmath}
\usepackage{amssymb}
\usepackage{graphicx}
\usepackage{bm}
\usepackage{multirow}
\usepackage{color}
\usepackage{mhchem}
\usepackage{braket}

\def\<{\langle}
\def\>{\rangle}
\def\ts{vdW$^{\rm TS}$}

\def\pbets{PBE+vdW$^{\rm TS}_{\rm SC}$}
\def\pbetsnsc{PBE+vdW$^{\rm TS}$}
\def\pbeots{PBE0+vdW$^{\rm TS}_{\rm SC}$}

\begin{document}
\title{Thermal Expansion in Dispersion-Bound Molecular Crystals}
\author{Hsin-Yu Ko}
\affiliation{Department of Chemistry, Princeton University, Princeton, NJ 08544, USA}
\author{Robert A. DiStasio Jr.}
\affiliation{Department of Chemistry and Chemical Biology, Cornell University, NY 14853, USA}
\author{Biswajit Santra}
\affiliation{Department of Chemistry, Princeton University, Princeton, NJ 08544, USA}
\author{Roberto Car}
\email{rcar@princeton.edu} 
\affiliation{Department of Chemistry, Princeton University, Princeton, NJ 08544, USA}
\affiliation{Department of Physics, Princeton University, Princeton, NJ 08544, USA}

\date{\today}

\begin{abstract}
We explore how anharmonicity, nuclear quantum effects (NQE), many-body dispersion interactions, and Pauli repulsion influence thermal properties of dispersion-bound molecular crystals. Accounting for anharmonicity with \textit{ab initio} molecular dynamics yields cell parameters accurate to within $2$\% of experiment for a set of pyridine-like molecular crystals at finite temperatures and pressures. From the experimental thermal expansion curve, we find that pyridine-I has a Debye temperature just above its melting point, indicating sizable NQE across the entire crystalline range of stability. We find that NQE lead to a substantial volume increase in pyridine-I ($\approx 40$\% more than classical thermal expansion at $153$~K) and attribute this to intermolecular Pauli repulsion promoted by intramolecular quantum fluctuations. When predicting delicate properties such as the thermal expansivity, we show that many-body dispersion interactions and sophisticated treatments of Pauli repulsion are needed in dispersion-bound molecular crystals.
\end{abstract}

\maketitle

Molecular crystals are versatile materials with widespread use across many fields~\cite{bernstein_polymorphism_2007,price_computational_2008}, including pharmaceuticals~\cite{almarsson_crystal_2004}, explosives~\cite{laurence_e_fried_design_2001}, and nonlinear optics~\cite{prakash_second_2006}. In these cases, properties such as biological activity of a drug, energy density of an explosive, and optical response of a nonlinear medium are all governed by the underlying structures of the molecular crystals and their (often numerous) polymorphs. This stresses the need for accurate and reliable theoretical methods for crystal structure prediction (CSP)~\cite{price_computational_2008,price_crystal_2008}, which not only provide key physical insight into such structure-property relationships, but also offer the promise of rational design of molecular crystals with novel and targeted properties~\cite{hoja_first-principles_2017}.

Despite the fact that all real-world solid-state applications occur at finite temperatures ($T$) and pressures ($p$), most CSP methods focus on determining structural properties (\textit{e.g.}, lattice parameters and cell volumes) at $0$~K. While such athermal predictions can be accurate for covalent and ionic solids, this approach is unlikely to provide quantitative structural information for non-covalently bound systems such as molecular crystals, which often have large thermal expansivities originating from relatively weak and highly anharmonic intermolecular forces. For example, the volume of the benzene molecular crystal increases by $2.7$\% from $4$~K--$138$~K~\cite{david_crystal_1992,bacon_crystallographic_1964}, while thermal effects in \ce{Si} are at least one order of magnitude smaller at similar temperatures~\cite{middelmann_thermal_2015}.

To predict how finite $T$ and $p$ influence structural properties in molecular crystals, one can utilize \textit{ab initio} molecular dynamics (AIMD)~\cite{marx_ab_2009} in the isobaric-isothermal ($NpT$) ensemble. In this technique, the quality of the predicted structures/properties is governed by the accuracy of the theoretical descriptions for the electrons and nuclei. With a quite favorable ratio of cost to accuracy, Density Functional Theory (DFT)~\cite{hohenberg_inhomogeneous_1964,kohn_self-consistent_1965} based on the generalized-gradient approximation (GGA) is often used to treat the electrons and has become the \textit{de facto} standard in first-principles simulations of condensed-phase systems in chemistry, physics, and materials science. Despite this widespread success, semi-local functionals cannot account for long-range dispersion or van der Waals (vdW) interactions, which are crucial for even qualitatively describing non-covalently bound molecular crystals~\cite{lu_ab_2009}. GGA-based functionals also suffer from spurious self-interaction error (SIE)~\cite{perdew_self-interaction_1981,cohen_insights_2008}, which leads to excessive delocalization of the molecular orbitals and charge densities. To account for non-bonded interactions, various corrections have been incorporated into DFT~\cite{klimes_perspective:_2012,grimme_dispersion-corrected_2016,hermann_first-principles_2017,berland_van_2015}, ranging from effective pairwise models~\cite{becke_exchange-hole_2007,tkatchenko_accurate_2009,grimme_consistent_2010,ferri_electronic_2015} and approaches that account for many-body dispersion interactions~\cite{tkatchenko_accurate_2012,distasio_jr._collective_2012,distasio_jr._many-body_2014,ambrosetti_long-range_2014,blood-forsythe_analytical_2016} to non-local functionals~\cite{dion_van_2004,vydrov_nonlocal_2009,lee_higher-accuracy_2010}. To ameliorate the SIE, hybrid functionals~\cite{becke_new_1993} incorporate a fraction of exact exchange in the DFT potential. Beyond the choice of functional, most AIMD simulations employ classical mechanics for the nuclear motion and neglect the quantum mechanical nature of the nuclei as they sample the potential energy surface (PES). Such nuclear quantum effects (NQE), \textit{e.g.}, zero-point motion and tunneling, can be accounted for using the Feynman path-integral (PI) approach~\cite{fosdick_numerical_1962,chandler_exploiting_1981,marx_ab_1996,tuckerman_efficient_1996,ceriotti_accelerating_2011,ceriotti_i-pi:_2014}.

\begin{figure}[ht!]
  \includegraphics[width=\linewidth]{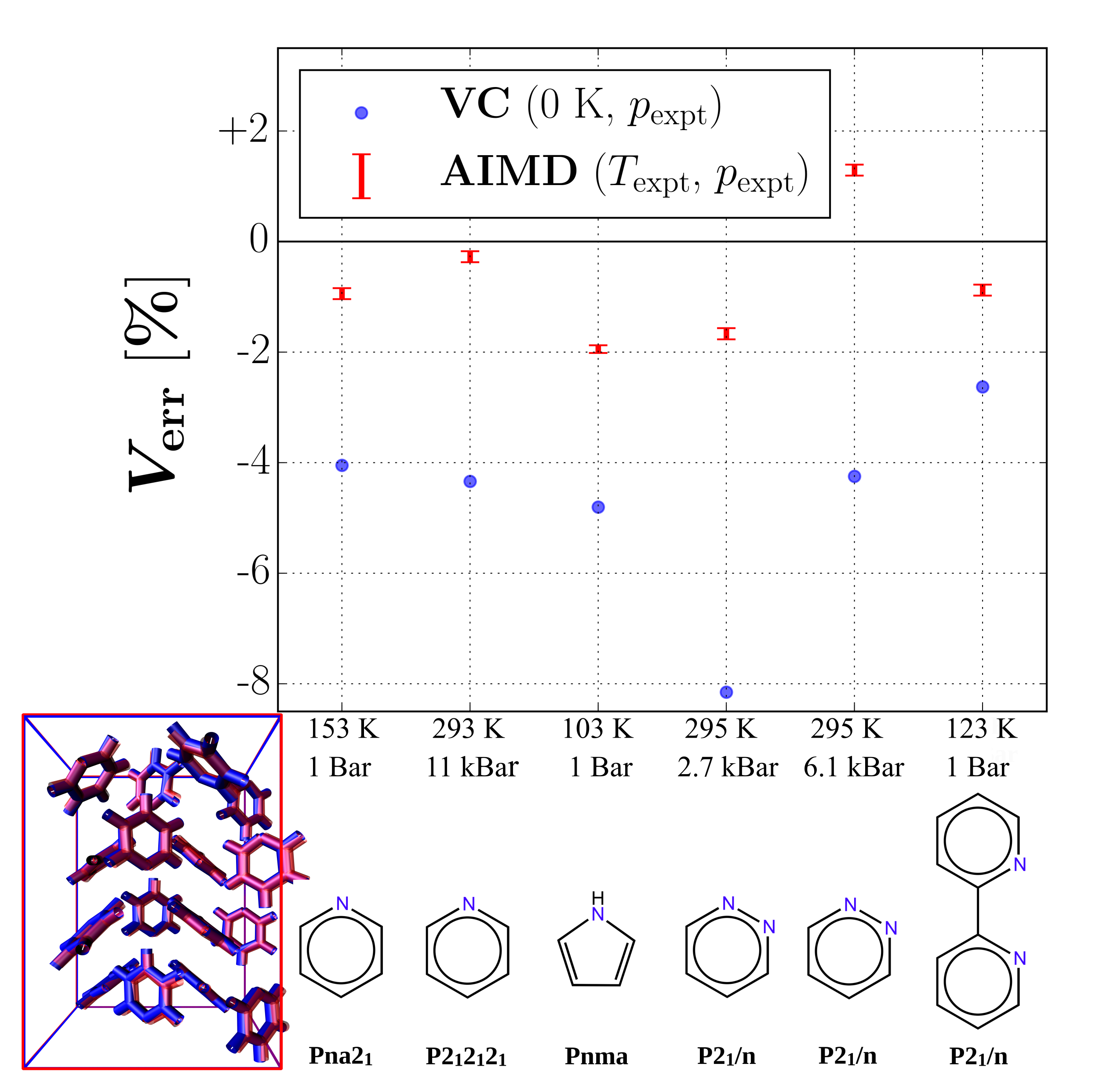}
  \caption{
  Predicted cell volumes ($V_{\rm pred}$) from VC optimizations and AIMD simulations using \pbets{} for the PLMC set. Errors are defined with respect to experiment ($V_{\rm expt}$) at the indicated thermodynamic conditions as $V_{\rm err}=(V_{\rm pred}-V_{\rm expt})/V_{\rm expt}$. \textit{Inset}: Overlay of predicted (blue) and experimental (red)~\cite{mootz_crystal_1981} pyridine-I structures.
  }
  \label{fig:overlay_n_otherPLMCs}
\end{figure}

In this Letter, we explore how anharmonicity, nuclear quantum fluctuations, many-body dispersion interactions, and Pauli repulsion influence structural and thermal properties in dispersion-bound molecular crystals at different thermodynamic conditions. As a first step, we investigate the influence of anharmonicity on the structural properties in a set of pyridine-like molecular crystals (PLMCs), comprised of the following N-heterocyclic aromatic compounds: pyridine (two polymorphs)~\cite{mootz_crystal_1981,crawford_isotopic_2009}, pyrrole~\cite{goddard_pyrrole_1997}, pyridazine (two different thermodynamic conditions)~\cite{podsiadlo_density_2010}, and bipyridine~\cite{kuhn_octahedral_2002} (Fig.~\ref{fig:overlay_n_otherPLMCs}). These molecules are pervasive throughout chemistry, biology, and agriculture~\cite{gilchrist_heterocyclic_2007} as common ligands and solvents, pharmacophores, and herbicide precursors. To quantify this influence on the PLMC cell parameters under experimental conditions ($T_{\rm expt}$, $p_{\rm expt}$), we performed variable-cell (VC) optimizations at ($0$~K, $p_{\rm expt}$) and $NpT$-based AIMD simulations at ($T_{\rm expt}$, $p_{\rm expt}$).

For this study, we employed the Perdew-Burke-Ernzerhof (PBE) GGA-based exchange-correlation (XC) functional~\cite{perdew_generalized_1996} in conjunction with a fully self-consistent (SC) implementation~\cite{distasio_jr._individual_2014,ferri_electronic_2015} of the Tkatchenko-Scheffler (TS) dispersion correction~\cite{tkatchenko_accurate_2009}, denoted by \pbets{} throughout. The \ts{} method is an effective pairwise ($C_6/R^6$) approach wherein all atomic parameters (\textit{e.g.}, dipole polarizabilities, vdW radii, and dispersion coefficients) are functionals of the electron density. This approach accounts for the unique chemical environment surrounding each atom and yields interatomic $C_6$ coefficients accurate to $\approx 5$\%~\cite{tkatchenko_accurate_2009,hermann_first-principles_2017}. When compared with low-$T$ experiments, VC optimizations with \pbetsnsc{} predict lattice parameters to $\approx 2$\% in crystals containing small organic molecules like ammonia, benzene, urea, and naphthalene~\cite{al-saidi_assessment_2012,bucko_tkatchenko-scheffler_2013,reilly_understanding_2013}. In the SC implementation, non-local correlation effects are accounted for in the charge density \textit{via} the dispersion contribution to the XC potential. Evaluation of the \pbets{} energy and forces ensures appropriate energy conservation during AIMD~\cite{distasio_jr._individual_2014} and can significantly affect binding energies in highly polarizable molecules and materials as well as coinage-metal work functions~\cite{ferri_electronic_2015}. The Car-Parrinello molecular dynamics (CPMD)~\cite{car_unified_1985} approach was used for all $NpT$ simulations in conjunction with massive Nos\'e-Hoover thermostat chains~\cite{tobias_molecular_1993} and the Parrinello-Rahman barostat~\cite{parrinello_crystal_1980}. All VC optimizations and CPMD simulations (for $\ge 10$~ps) were performed using \textsc{Quantum ESPRESSO} (QE)~\cite{giannozzi_quantum_2009,giannozzi_advanced_2017} at a constant (planewave) kinetic energy cutoff following Ref.~\cite{bernasconi_first-principle-constant_1995} to avoid Pulay-like stress from cell fluctuations~\cite{sup-all}.

Fig.~\ref{fig:overlay_n_otherPLMCs} compares the predicted volumes from VC optimizations and AIMD simulations with experiment, clearly demonstrating that anharmonicity effects are indeed non-negligible in determining this structural property. VC optimizations always underestimate this quantity and the inclusion of anharmonicity \textit{via} $NpT$-based AIMD systematically reduces the mean absolute error (MAE) from $4.7$\% to $1.2$\% in the predicted volumes. In fact, the influence of anharmonicity can be quite substantial in the PLMC set, as evidenced by the $6.4$\% change in $V_{\rm err}$ for pyridazine at ($295$~K, $2.7$~kBar). We note that the extent to which anharmonicity will influence cell volume expansion depends on a complex interplay between $p_{\rm expt}$ and the cohesive forces at work in the crystal (which act together to suppress expansion) and $T_{\rm expt}$ (which provides thermal energy for PES exploration).

AIMD simulations also yield PLMC lattice parameters that agree remarkably well with experiment (Table~\ref{tab:npt_cel_abc} and Table~S1). By accounting for anharmonicity, AIMD systematically reduce the MAE in the predicted lattice parameters from $2.0$\% to $1.3$\% with respect to experiment. As seen above, VC optimizations tend to underestimate PLMC lattice parameters; however, this trend does not always hold as evidenced by the slight \textit{negative} linear thermal expansion observed along the $c$ axis in pyridine-II. This predicted effect is consistent with the experimental data~\cite{crawford_isotopic_2009} and reproduces the reference lattice parameter with extremely high fidelity. By considering the lattice parameter fluctuations throughout the AIMD trajectory, we found that the $c$ axis was not the softest (most flexible) dimension in pyridine-II, hence the apparent negative linear thermal expansion in this molecular crystal has a distinctly different origin than that of methanol monohydrate~\cite{fortes_negative_2011}. Since this effect is also observed during GGA-based AIMD (which do not account for dispersion interactions), this phenomenon is most likely electrostatic in nature for pyridine-II. In addition, the structure and orientation of the individual molecules inside the PLMC unit cells are also well described by AIMD with \pbets{} (Fig.~\ref{fig:overlay_n_otherPLMCs} and Fig.~S1), with associated root-mean-square deviations (RMSD) of $0.17$ \AA\ across this set of dispersion-bound molecular crystals.

\begin{table}[t!]
  \caption{ 
  Predicted and experimental structural properties for the pyridine-I and pyridine-II molecular crystals. All simulations were performed using \pbets{} and the numbers in parentheses denote uncertainties in the predicted values.}
  \begin{tabular}{c | c c c c}
    \hline\hline
    Pyridine-I                    & $a$ [\AA]    & $b$  [\AA]  &   $c$ [\AA]  & $V_{\rm err}$ (\%)  \\
    \hline
    VC ($0$~K, $1$~Bar)               & $17.25$      & $8.88$      &   $11.14$    & $-4.0$    \\
    AIMD ($153$~K, $1$~Bar)           & $17.43(3)$   & $8.92(2)$   &   $11.31(5)$ & $-0.9(1)$ \\
    PI-AIMD ($153$~K, $1$~Bar)        & $17.51(4)$   & $8.95(3)$   &  $11.44(6)$  & $+0.3(1)$ \\
    Expt.~\cite{mootz_crystal_1981} & $17.52$      & $8.97$      &   $11.35$    & --        \\
    \hline
    \multicolumn{5}{c}{\vspace{-9pt}} \\
    \hline
    Pyridine-II                   & $a$ [\AA]    & $b$  [\AA]  &   $c$ [\AA]  & $V_{\rm err}$   (\%) \\
    \hline
    VC ($0$~K, $11$~kBar)             &  $5.33$      &  $6.56$     &   $11.30$    &  $-4.3$    \\
    AIMD ($298$~K, $11$~kBar)         &  $5.46(1)$   &  $6.72(4)$  &   $11.23(5)$ &  $-0.4(1)$ \\ 
    Expt.~\cite{crawford_isotopic_2009} &  $5.40$      &  $6.80$     &   $11.23$    &   --       \\
    \hline\hline
  \end{tabular}
  \label{tab:npt_cel_abc}
\end{table}

Based on these findings, we conclude that structural predictions are significantly improved when anharmonicity is accounted for \textit{via} $NpT$-based AIMD simulations, yielding \textit{finite-temperature} structural properties in dispersion-bound molecular crystals that are within $2$\% of experiment. However, the results reported herein still systematically underestimate the experimental PLMC cell volumes. For more accurate and reliable predictions, we find that NQE (such as zero-point fluctuations), many-body dispersion interactions, and Pauli repulsion all have a non-negligible influence over the structural and thermal properties of dispersion-bound molecular crystals. To demonstrate this, we now focus our attention on a detailed case study of the pyridine-I polymorph.

\begin{figure}[t!]
  \includegraphics[width=\linewidth]{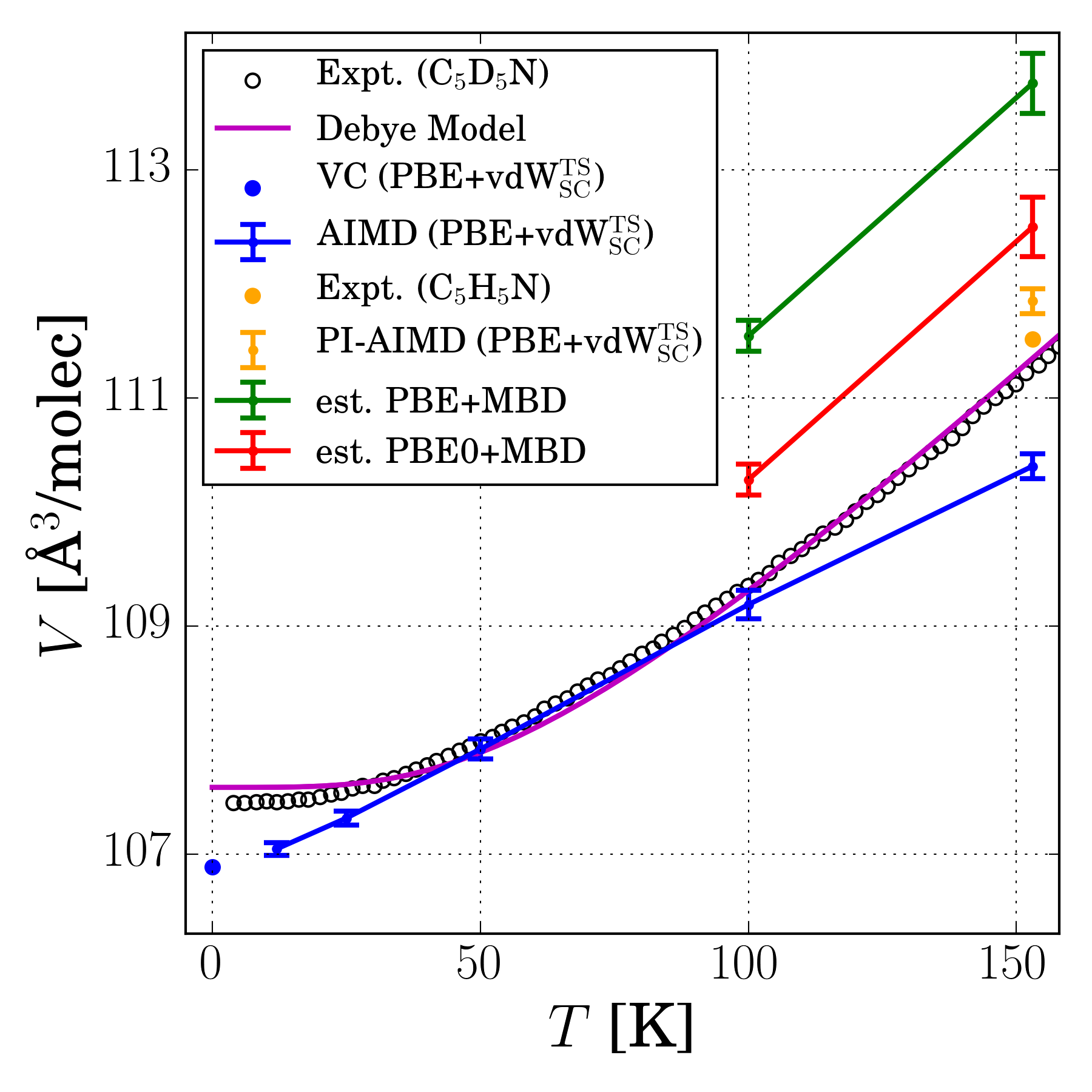}
  \caption{
  Predicted and experimental thermal expansion curves for pyridine-I. Experimental data is included for pyridine-I (\ce{C5H5N}, gold circle), from single-crystal X-ray diffraction~\cite{mootz_crystal_1981}, and fully deuterated pyridine-I (\ce{C5D5N}, open black circles), from neutron powder diffraction~\cite{crawford_isotopic_2009}. A fit of the experimental thermal expansion curve for deuterated pyridine-I using the Debye model for $V(T)$ is given by the purple line (Eq.~\eqref{eq:debye_v}). Theoretical data is included for VC optimizations (blue circle), AIMD simulations (blue line), and PI-AIMD simulations (gold circle with error bar) at the \pbets{} level; estimated PBE+MBD results (green line, Eq.~\eqref{eq:reweigh}); estimated PBE0+MBD results (red line, see text for details).
  }
  \label{fig:exp}
\end{figure}

While AIMD simulations are able to furnish accurate structural properties for the PLMCs across a range of thermodynamic conditions, the shape of the thermal expansion curve for deuterated pyridine-I from neutron powder diffraction experiments~\cite{crawford_isotopic_2009} significantly differs from our theoretical predictions (Fig.~\ref{fig:exp}). In this regard, the predicted $V(T)$ curve is linear across the entire $T$ range considered (\textit{i.e.}, $12$~K--$153$~K at $p_{\rm expt}=1$~Bar), reflecting the use of classical mechanics for the nuclear motion. The experimental curve, on the other hand, shows non-linear behavior in this $T$ interval, with significant deviations from linearity at low temperatures, \textit{i.e.}, for $T \le 50$~K. This observation strongly indicates that NQE (in particular zero-point motion) play a non-negligible role in governing the structural and thermal properties of this dispersion-bound molecular crystal.

To gain further insight into the thermal expansion behavior in this system, we utilize the Debye model, which is an isotropic acoustic approximation for the phonons in a solid. Within this framework, $V(T)$ can be derived from the corresponding Gibbs free energy (at a given $p$) as~\cite{sup-all}:
\begin{equation}
  V(T) = V(0) + \left[ 3N k_{B} \frac{\Theta_D'}{\Theta_D} \, \mathbf{D}\!\left( \frac{\Theta_D}{T} \right) \right] T ,
  \label{eq:debye_v}
\end{equation}
in which $V(0)$ is the cell volume at $0$~K, $N$ is the number of atoms, $\Theta_D=\Theta_D(p)$ is the Debye temperature, $\Theta_D'=\text{d} \, \Theta_D(p)/\text{d} \, p$ is the pressure derivative of $\Theta_D$ (which accounts for anharmonicity in the underlying PES), and $\mathbf{D}(\boldsymbol{\cdot})$ is the Debye function~\cite{landau_statistical_1996}. Quite interestingly, we find that the experimental thermal expansion curve for \ce{C5D5N} can be fit rather well with Eq.~\eqref{eq:debye_v}, as shown by the purple line in Fig.~\ref{fig:exp}. A similarly good fit using the Debye interpolation formula was obtained for the methanol monohydrate molecular crystal~\cite{fortes_negative_2011}. The validity of the Debye model for thermal expansion in pyridine-I is further supported by the physical value for the Debye temperature obtained from the fit, namely, $\Theta_D=235(5)$~K. This corresponds to an average sound velocity of $1710$~m/s in this system, which falls within the experimentally determined range for the sound velocity of the closely related benzene molecular crystal~\cite{heseltine_thesis_determination_1962,sup-all}.

The fact that $\Theta_D$ is slightly above the melting temperature of pyridine-I ($T_m = 232$~K) suggests that NQE should have a sizable influence across the entire crystalline range of stability in this polymorph. To directly confirm the importance of NQE in determining the structure of pyridine-I, we performed a PI-AIMD simulation using \pbets{} at ($153$~K, $1$~Bar)~\cite{ceriotti_nuclear_2009,ceriotti_accelerating_2011,ceriotti_i-pi:_2014,sup-all}. When compared to the $3$\% volume expansion due to classical thermal fluctuations (\textit{cf}. the difference between the VC optimization at $0$~K and AIMD simulation at $153$~K, see Table~\ref{tab:npt_cel_abc}), we find that the inclusion of NQE results in an additional $1.2$\% expansion in the cell volume. This change is quite sizable ($\approx 40$\% of the classical thermal expansion) and further reduces $V_{\rm err}$ in pyridine-I to $+0.3$\% with respect to experiment. This volume expansion is primarily attributed to repulsive \textit{intermolecular} contacts in the pyridine-I molecular crystal that have been induced by \textit{intramolecular} quantum fluctuations. In this regard, the probability of finding \ce{H} atoms on neighboring pyridine molecules at distances less than the sum of their vdW radii increases by nearly $30$\% when NQE are accounted for \textit{via} PI-AIMD simulations~\cite{sup-all}. Taken together, all of these findings demonstrate that NQE play a substantive role in governing the structural properties of this dispersion-bound molecular crystal.

Considering now the thermal expansivity (or thermal expansion coefficient),
\begin{equation}
  \alpha (T) = \frac{1}{V(T)} \left( \frac{\partial \, V(T)}{\partial \, T} \right)_p ,
  \label{eq:expansivity}
\end{equation}
we determined an experimental value of $\alpha = 3.5 \times 10^{-4}\text{ K}^{-1}$ for pyridine-I at ($153$~K, $1$~Bar) based on the \ce{C5D5N} thermal expansion curve~\cite{crawford_isotopic_2009}. This value agrees quite well with the analytical finding from the Debye interpolation, \textit{i.e.}, $\alpha = 3.7 \times 10^{-4}\text{ K}^{-1}$, further illustrating the utility of this model in describing this system. However, the $\alpha$ value from classical AIMD simulations using \pbets{} ($\alpha = 2.1 \times 10^{-4}\text{ K}^{-1}$) significantly underestimates the experimental value by $\approx 40$\%. Since cohesion in pyridine-I is dominated by dispersion interactions~\cite{sup-all}, this suggests that \pbets{} overestimates the cohesive forces at work in this non-covalently bound molecular crystal. This finding is consistent with other studies on molecular crystal lattice energies with this method~\cite{reilly_understanding_2013}. As such, we now investigate how a more comprehensive treatment of the beyond-pairwise many-body dispersion forces impacts our prediction of this thermal property in pyridine-I.

\begin{table}[t!]
  \caption{ 
  Thermal expansivity ($\alpha$) values for pyridine-I at ($153$~K, $1$~Bar) from theoretical simulations (at the \pbets{}, est. PBE+MBD, and est. PBE0+MBD levels), the Debye model, and experiment. Errors are reported with respect to the experimental value and the numbers in parentheses denote uncertainties in $\alpha$. 
  }
  \begin{tabular}{c | c | c}
    \hline
    \hline
    Pyridine-I & $\alpha$ [$10^{-4}$~K$^{-1}$] &  $\alpha_{\rm err}$ (\%) \\
    \hline
    \pbets{}       & $2.1(3)$  & $-40.0$ \\
    est. PBE+MBD   & $3.7(5)$  & $+5.7$ \\
    est. PBE0+MBD  & $3.7(5)$  & $+5.7$ \\
    Debye Model    & $3.65(4)$ & $+4.3$ \\
    Expt.~\cite{crawford_isotopic_2009} & $3.5(1)$ & -- \\
    \hline
    \hline
  \end{tabular}
  \label{tab:thermal_expansivity}
\end{table}

Beyond-pairwise dispersion interactions include terms such as the three-body Axilrod-Teller-Muto (ATM) contribution ($C_9/R^9$)~\cite{axilrod_interaction_1943,muto_force_1943}, which is more short-ranged than the $C_6/R^6$ term in the effective-pairwise \ts{} level and often provides a repulsive contribution to the binding energy. Since the inclusion of the ATM term alone is usually not sufficient to describe the full many-body expansion of the dispersion energy~\cite{ambrosetti_hard_2014}, we employ the many body dispersion (MBD) model~\cite{tkatchenko_accurate_2012,distasio_jr._collective_2012,distasio_jr._many-body_2014,ambrosetti_long-range_2014,blood-forsythe_analytical_2016} to investigate how these higher-order non-bonded interactions affect the structural and thermal properties in pyridine-I. The MBD approach furnishes a description of all $N$-body dispersion energy contributions by mapping the atoms comprising the system onto a set of coupled quantum harmonic oscillators, and then computing the long-range correlation energy in the random-phase approximation (RPA)~\cite{tkatchenko_interatomic_2013,distasio_jr._many-body_2014,ambrosetti_long-range_2014}. When coupled with DFT, MBD has been shown to provide an accurate and reliable description of the non-covalent interactions in molecules and materials~\cite{hermann_first-principles_2017}, ranging from molecular crystals~\cite{reilly_understanding_2013,marom_many-body_2013,reilly_report_2016} to complex polarizable nanostructures~\cite{ambrosetti_wavelike_2016,ambrosetti_physical_2017}.

To account for many-body dispersion interactions, we estimated the average cell volume at the PBE+MBD level ($\braket{V}_{\rm MBD}$) by Boltzmann reweighting the configurations from the \pbets{} trajectory, \textit{i.e.}, 
\begin{equation}
  \braket{V}_{\rm MBD} = \frac{\braket{V \exp \left[ -\beta \left( U_{\rm MBD}-U_{\rm TS} \right) \right] }_{\rm TS}}{\braket{\exp \left[ -\beta \left( U_{\rm MBD} - U_{\rm TS} \right) \right] }_{\rm TS}} ,
  \label{eq:reweigh}
\end{equation}
in which $\beta$ is the inverse temperature, $U_{\rm TS}$ and $U_{\rm MBD}$ are the corresponding dispersion energies from these two methods, and $\braket{\,\cdot\,}_{\rm TS}$ represents a statistical average over the \pbets{} ensemble~\cite{sup-all}. The resulting estimates for $\braket{V}_{\rm MBD}$ are shown in Fig.~\ref{fig:exp} and were used to determine that $\alpha = 3.7 \times 10^{-4}$~K$^{-1}$ at the PBE+MBD level, which is in significantly better agreement with the experimental value than \pbets{} (Table~\ref{tab:thermal_expansivity}). However, the estimated PBE+MBD cell volumes are noticeably larger than experiment, with predictions that are now less accurate than \pbets{}. Since MBD provides a more comprehensive treatment of dispersion interactions, this effect is likely a manifestation of other deficiencies present in the XC functional. In particular, the semi-local PBE functional suffers from SIE and therefore provides an inaccurate description of the Pauli repulsion in this system, the effect of which will be considered in detail below.

Repulsive intermolecular interactions in this molecular crystal mainly originate from overlapping electron clouds on neighboring pyridine molecules. Due to the presence of SIE, molecular charge densities at the PBE level tend to be too diffuse~\cite{cohen_insights_2008}, which leads to increased density overlap and thus a substantial overestimate of the Pauli repulsion. This error can be largely ameliorated by including a fraction of exact exchange, as accomplished by hybrid DFT functionals like PBE0~\cite{perdew_rationale_1996}. Here, we estimate the extent of this effect by the difference between cell volumes obtained from VC optimizations using \pbeots{}~\cite{wu_order-n_2009,distasio_jr._individual_2014,exx-draft} and \pbets{}. With $\Delta V = -1.02$ \AA{}$^3$/molec, we estimate the PBE0+MBD volume by adding this constant shift to the PBE+MBD results above (Fig.~\ref{fig:exp}). This largely corrects the overestimation of the cell volume with PBE+MBD and the resulting estimated PBE0+MBD volumes are now in excellent agreement with both the experimental volume (on an absolute scale) and thermal expansivity. We stress here that an improved theoretical description of the Pauli repulsion will be of particular importance when coupled with NQE, which substantially increase the amount of vdW overlap in this molecular crystal.

In this Letter, we explored how a complex interplay between anharmonicity, NQE, many-body dispersion interactions, and Pauli repulsion influence the structural and thermal properties of dispersion-bound molecular crystals. By focusing on pyridine-I, we showed that the Debye model is well-suited to describe the thermal expansion behavior in this system across the entire range of experimental temperatures. With a Debye temperature just above the melting point, we expect that NQE will be sizable across the entire crystalline range of stability in this polymorph. At low $T$, PI-AIMD simulations become computationally intractable (due to the steep increase in the Trotter dimension) and it would be more efficient to include NQE \textit{via} the quasiharmonic or self-consistent harmonic approximations~\cite{hooton_li._1955,errea_first-principles_2013,errea_anharmonic_2014}. Beyond the structural and thermal properties considered herein, the existence of thermodynamically relevant polymorphs further advocates for the determination of structures, stabilities, and properties of molecular crystals under $NpT$ conditions. Based on the findings presented in this work, free energy calculations that simultaneously account for nuclear quantum fluctuations, many-body dispersion interactions, and a sophisticated treatment of Pauli repulsion will be required for an accurate and reliable description of dispersion-bound molecular crystals.

All authors gratefully acknowledge support from the U.S. Department of Energy under Grant Nos. DE-SC0005180 and DE-SC0008626. R.D. acknowledges partial support from Cornell University through start-up funding and the Cornell Center for Materials Research (CCMR) with funding from the National Science Foundation MRSEC program (DMR-1719875). This research used resources of the National Energy Research Scientific Computing (NERSC) Center, which is supported by the Office of Science of the U.S. Department of Energy under Contract No. DE-AC02-05CH11231. This research used resources of the Argonne Leadership Computing Facility at Argonne National Laboratory, which is supported by the Office of Science of the U.S. Department of Energy under Contract No. DE-AC02-06CH11357. Additional resources were provided by the Terascale Infrastructure for Groundbreaking Research in Science and Engineering (TIGRESS) High Performance Computing Center and Visualization Laboratory at Princeton University.


\begin{thebibliography}{75}
\expandafter\ifx\csname natexlab\endcsname\relax\def\natexlab#1{#1}\fi
\expandafter\ifx\csname bibnamefont\endcsname\relax
  \def\bibnamefont#1{#1}\fi
\expandafter\ifx\csname bibfnamefont\endcsname\relax
  \def\bibfnamefont#1{#1}\fi
\expandafter\ifx\csname citenamefont\endcsname\relax
  \def\citenamefont#1{#1}\fi
\expandafter\ifx\csname url\endcsname\relax
  \def\url#1{\texttt{#1}}\fi
\expandafter\ifx\csname urlprefix\endcsname\relax\def\urlprefix{URL }\fi
\providecommand{\bibinfo}[2]{#2}
\providecommand{\eprint}[2][]{\url{#2}}

\bibitem[{\citenamefont{Bernstein}(2007)}]{bernstein_polymorphism_2007}
\bibinfo{author}{\bibfnamefont{J.}~\bibnamefont{Bernstein}},
  \emph{\bibinfo{title}{Polymorphism in {Molecular} {Crystals}}}
  (\bibinfo{publisher}{Oxford University Press}, \bibinfo{address}{New York},
  \bibinfo{year}{2007}).

\bibitem[{\citenamefont{Price}(2008{\natexlab{a}})}]{price_computational_2008}
\bibinfo{author}{\bibfnamefont{S.~L.} \bibnamefont{Price}},
  \bibinfo{journal}{Int. Rev. Phys. Chem.} \textbf{\bibinfo{volume}{27}},
  \bibinfo{pages}{541} (\bibinfo{year}{2008}{\natexlab{a}}).

\bibitem[{\citenamefont{Almarsson and
  Zaworotko}(2004)}]{almarsson_crystal_2004}
\bibinfo{author}{\bibfnamefont{{\"O}.}~\bibnamefont{Almarsson}}
  \bibnamefont{and} \bibinfo{author}{\bibfnamefont{M.~J.}
  \bibnamefont{Zaworotko}}, \bibinfo{journal}{Chem. Commun.}
  \textbf{\bibinfo{volume}{0}}, \bibinfo{pages}{1889} (\bibinfo{year}{2004}).

\bibitem[{\citenamefont{Fried et~al.}(2001)\citenamefont{Fried, Manaa, Pagoria,
  and Simpson}}]{laurence_e_fried_design_2001}
\bibinfo{author}{\bibfnamefont{L.~E.} \bibnamefont{Fried}},
  \bibinfo{author}{\bibfnamefont{M.~R.} \bibnamefont{Manaa}},
  \bibinfo{author}{\bibfnamefont{P.~F.} \bibnamefont{Pagoria}},
  \bibnamefont{and} \bibinfo{author}{\bibfnamefont{R.~L.}
  \bibnamefont{Simpson}}, \bibinfo{journal}{Annu. Rev. Mater. Sci.}
  \textbf{\bibinfo{volume}{31}}, \bibinfo{pages}{291} (\bibinfo{year}{2001}).

\bibitem[{\citenamefont{Prakash and Radhakrishnan}(2006)}]{prakash_second_2006}
\bibinfo{author}{\bibfnamefont{M.~J.} \bibnamefont{Prakash}} \bibnamefont{and}
  \bibinfo{author}{\bibfnamefont{T.~P.} \bibnamefont{Radhakrishnan}},
  \bibinfo{journal}{Chem. Mater.} \textbf{\bibinfo{volume}{18}},
  \bibinfo{pages}{2943} (\bibinfo{year}{2006}).

\bibitem[{\citenamefont{Price}(2008{\natexlab{b}})}]{price_crystal_2008}
\bibinfo{author}{\bibfnamefont{S.~L.} \bibnamefont{Price}},
  \bibinfo{journal}{Phys. Chem. Chem. Phys.} \textbf{\bibinfo{volume}{10}},
  \bibinfo{pages}{1996} (\bibinfo{year}{2008}{\natexlab{b}}).

\bibitem[{\citenamefont{Hoja et~al.}(2017)\citenamefont{Hoja, Reilly, and
  Tkatchenko}}]{hoja_first-principles_2017}
\bibinfo{author}{\bibfnamefont{J.}~\bibnamefont{Hoja}},
  \bibinfo{author}{\bibfnamefont{A.~M.} \bibnamefont{Reilly}},
  \bibnamefont{and}
  \bibinfo{author}{\bibfnamefont{A.}~\bibnamefont{Tkatchenko}},
  \bibinfo{journal}{WIREs Comput Mol Sci} \textbf{\bibinfo{volume}{7}},
  \bibinfo{pages}{1294} (\bibinfo{year}{2017}).

\bibitem[{\citenamefont{David et~al.}(1992)\citenamefont{David, Ibberson,
  Jeffrey, and Ruble}}]{david_crystal_1992}
\bibinfo{author}{\bibfnamefont{W.~I.~F.} \bibnamefont{David}},
  \bibinfo{author}{\bibfnamefont{R.~M.} \bibnamefont{Ibberson}},
  \bibinfo{author}{\bibfnamefont{G.~A.} \bibnamefont{Jeffrey}},
  \bibnamefont{and} \bibinfo{author}{\bibfnamefont{J.~R.} \bibnamefont{Ruble}},
  \bibinfo{journal}{Physica B (Amsterdam, Neth.)}
  \textbf{\bibinfo{volume}{180-181}}, \bibinfo{pages}{597}
  (\bibinfo{year}{1992}).

\bibitem[{\citenamefont{Bacon et~al.}(1964)\citenamefont{Bacon, Curry, and
  Wilson}}]{bacon_crystallographic_1964}
\bibinfo{author}{\bibfnamefont{G.~E.} \bibnamefont{Bacon}},
  \bibinfo{author}{\bibfnamefont{N.~A.} \bibnamefont{Curry}}, \bibnamefont{and}
  \bibinfo{author}{\bibfnamefont{S.~A.} \bibnamefont{Wilson}},
  \bibinfo{journal}{Proc. R. Soc. Lond. A} \textbf{\bibinfo{volume}{279}},
  \bibinfo{pages}{98} (\bibinfo{year}{1964}).

\bibitem[{\citenamefont{Middelmann et~al.}(2015)\citenamefont{Middelmann,
  Walkov, Bartl, and Sch{\"o}del}}]{middelmann_thermal_2015}
\bibinfo{author}{\bibfnamefont{T.}~\bibnamefont{Middelmann}},
  \bibinfo{author}{\bibfnamefont{A.}~\bibnamefont{Walkov}},
  \bibinfo{author}{\bibfnamefont{G.}~\bibnamefont{Bartl}}, \bibnamefont{and}
  \bibinfo{author}{\bibfnamefont{R.}~\bibnamefont{Sch{\"o}del}},
  \bibinfo{journal}{Phys. Rev. B} \textbf{\bibinfo{volume}{92}},
  \bibinfo{pages}{174113} (\bibinfo{year}{2015}).

\bibitem[{\citenamefont{Marx and Hutter}(2009)}]{marx_ab_2009}
\bibinfo{author}{\bibfnamefont{D.}~\bibnamefont{Marx}} \bibnamefont{and}
  \bibinfo{author}{\bibfnamefont{J.}~\bibnamefont{Hutter}},
  \emph{\bibinfo{title}{Ab {Initio} {Molecular} {Dynamics}: {Basic} {Theory}
  and {Advanced} {Methods}}} (\bibinfo{publisher}{Cambridge University Press},
  \bibinfo{address}{Cambridge}, \bibinfo{year}{2009}).

\bibitem[{\citenamefont{Hohenberg and
  Kohn}(1964)}]{hohenberg_inhomogeneous_1964}
\bibinfo{author}{\bibfnamefont{P.}~\bibnamefont{Hohenberg}} \bibnamefont{and}
  \bibinfo{author}{\bibfnamefont{W.}~\bibnamefont{Kohn}},
  \bibinfo{journal}{Phys. Rev.} \textbf{\bibinfo{volume}{136}},
  \bibinfo{pages}{B864} (\bibinfo{year}{1964}).

\bibitem[{\citenamefont{Kohn and Sham}(1965)}]{kohn_self-consistent_1965}
\bibinfo{author}{\bibfnamefont{W.}~\bibnamefont{Kohn}} \bibnamefont{and}
  \bibinfo{author}{\bibfnamefont{L.~J.} \bibnamefont{Sham}},
  \bibinfo{journal}{Phys. Rev.} \textbf{\bibinfo{volume}{140}},
  \bibinfo{pages}{A1133} (\bibinfo{year}{1965}).

\bibitem[{\citenamefont{Lu et~al.}(2009)\citenamefont{Lu, Li, Rocca, and
  Galli}}]{lu_ab_2009}
\bibinfo{author}{\bibfnamefont{D.}~\bibnamefont{Lu}},
  \bibinfo{author}{\bibfnamefont{Y.}~\bibnamefont{Li}},
  \bibinfo{author}{\bibfnamefont{D.}~\bibnamefont{Rocca}}, \bibnamefont{and}
  \bibinfo{author}{\bibfnamefont{G.}~\bibnamefont{Galli}},
  \bibinfo{journal}{Phys. Rev. Lett.} \textbf{\bibinfo{volume}{102}},
  \bibinfo{pages}{206411} (\bibinfo{year}{2009}).

\bibitem[{\citenamefont{Perdew and
  Zunger}(1981)}]{perdew_self-interaction_1981}
\bibinfo{author}{\bibfnamefont{J.~P.} \bibnamefont{Perdew}} \bibnamefont{and}
  \bibinfo{author}{\bibfnamefont{A.}~\bibnamefont{Zunger}},
  \bibinfo{journal}{Phys. Rev. B} \textbf{\bibinfo{volume}{23}},
  \bibinfo{pages}{5048} (\bibinfo{year}{1981}).

\bibitem[{\citenamefont{Cohen et~al.}(2008)\citenamefont{Cohen,
  Mori-S{\'a}nchez, and Yang}}]{cohen_insights_2008}
\bibinfo{author}{\bibfnamefont{A.~J.} \bibnamefont{Cohen}},
  \bibinfo{author}{\bibfnamefont{P.}~\bibnamefont{Mori-S{\'a}nchez}},
  \bibnamefont{and} \bibinfo{author}{\bibfnamefont{W.}~\bibnamefont{Yang}},
  \bibinfo{journal}{Science} \textbf{\bibinfo{volume}{321}},
  \bibinfo{pages}{792} (\bibinfo{year}{2008}).

\bibitem[{\citenamefont{Klime{\v s} and
  Michaelides}(2012)}]{klimes_perspective:_2012}
\bibinfo{author}{\bibfnamefont{J.}~\bibnamefont{Klime{\v s}}} \bibnamefont{and}
  \bibinfo{author}{\bibfnamefont{A.}~\bibnamefont{Michaelides}},
  \bibinfo{journal}{J. Chem. Phys.} \textbf{\bibinfo{volume}{137}},
  \bibinfo{pages}{120901} (\bibinfo{year}{2012}).

\bibitem[{\citenamefont{Grimme et~al.}(2016)\citenamefont{Grimme, Hansen,
  Brandenburg, and Bannwarth}}]{grimme_dispersion-corrected_2016}
\bibinfo{author}{\bibfnamefont{S.}~\bibnamefont{Grimme}},
  \bibinfo{author}{\bibfnamefont{A.}~\bibnamefont{Hansen}},
  \bibinfo{author}{\bibfnamefont{J.~G.} \bibnamefont{Brandenburg}},
  \bibnamefont{and}
  \bibinfo{author}{\bibfnamefont{C.}~\bibnamefont{Bannwarth}},
  \bibinfo{journal}{Chem. Rev.} \textbf{\bibinfo{volume}{116}},
  \bibinfo{pages}{5105} (\bibinfo{year}{2016}).

\bibitem[{\citenamefont{Hermann et~al.}(2017)\citenamefont{Hermann,
  DiStasio~Jr., and Tkatchenko}}]{hermann_first-principles_2017}
\bibinfo{author}{\bibfnamefont{J.}~\bibnamefont{Hermann}},
  \bibinfo{author}{\bibfnamefont{R.~A.} \bibnamefont{DiStasio~Jr.}},
  \bibnamefont{and}
  \bibinfo{author}{\bibfnamefont{A.}~\bibnamefont{Tkatchenko}},
  \bibinfo{journal}{Chem. Rev.} \textbf{\bibinfo{volume}{117}},
  \bibinfo{pages}{4714} (\bibinfo{year}{2017}).

\bibitem[{\citenamefont{Berland et~al.}(2015)\citenamefont{Berland, Cooper,
  Lee, Schr{\"o}der, Thonhauser, Hyldgaard, and Lundqvist}}]{berland_van_2015}
\bibinfo{author}{\bibfnamefont{K.}~\bibnamefont{Berland}},
  \bibinfo{author}{\bibfnamefont{V.~R.} \bibnamefont{Cooper}},
  \bibinfo{author}{\bibfnamefont{K.}~\bibnamefont{Lee}},
  \bibinfo{author}{\bibfnamefont{E.}~\bibnamefont{Schr{\"o}der}},
  \bibinfo{author}{\bibfnamefont{T.}~\bibnamefont{Thonhauser}},
  \bibinfo{author}{\bibfnamefont{P.}~\bibnamefont{Hyldgaard}},
  \bibnamefont{and} \bibinfo{author}{\bibfnamefont{B.~I.}
  \bibnamefont{Lundqvist}}, \bibinfo{journal}{Rep. Prog. Phys.}
  \textbf{\bibinfo{volume}{78}}, \bibinfo{pages}{066501}
  (\bibinfo{year}{2015}).

\bibitem[{\citenamefont{Becke and Johnson}(2007)}]{becke_exchange-hole_2007}
\bibinfo{author}{\bibfnamefont{A.~D.} \bibnamefont{Becke}} \bibnamefont{and}
  \bibinfo{author}{\bibfnamefont{E.~R.} \bibnamefont{Johnson}},
  \bibinfo{journal}{J. Chem. Phys.} \textbf{\bibinfo{volume}{127}},
  \bibinfo{pages}{154108} (\bibinfo{year}{2007}).

\bibitem[{\citenamefont{Tkatchenko and
  Scheffler}(2009)}]{tkatchenko_accurate_2009}
\bibinfo{author}{\bibfnamefont{A.}~\bibnamefont{Tkatchenko}} \bibnamefont{and}
  \bibinfo{author}{\bibfnamefont{M.}~\bibnamefont{Scheffler}},
  \bibinfo{journal}{Phys. Rev. Lett.} \textbf{\bibinfo{volume}{102}},
  \bibinfo{pages}{073005} (\bibinfo{year}{2009}).

\bibitem[{\citenamefont{Grimme et~al.}(2010)\citenamefont{Grimme, Antony,
  Ehrlich, and Krieg}}]{grimme_consistent_2010}
\bibinfo{author}{\bibfnamefont{S.}~\bibnamefont{Grimme}},
  \bibinfo{author}{\bibfnamefont{J.}~\bibnamefont{Antony}},
  \bibinfo{author}{\bibfnamefont{S.}~\bibnamefont{Ehrlich}}, \bibnamefont{and}
  \bibinfo{author}{\bibfnamefont{H.}~\bibnamefont{Krieg}}, \bibinfo{journal}{J.
  Chem. Phys.} \textbf{\bibinfo{volume}{132}}, \bibinfo{pages}{154104}
  (\bibinfo{year}{2010}).

\bibitem[{\citenamefont{Ferri et~al.}(2015)\citenamefont{Ferri, DiStasio~Jr.,
  Ambrosetti, Car, and Tkatchenko}}]{ferri_electronic_2015}
\bibinfo{author}{\bibfnamefont{N.}~\bibnamefont{Ferri}},
  \bibinfo{author}{\bibfnamefont{R.~A.} \bibnamefont{DiStasio~Jr.}},
  \bibinfo{author}{\bibfnamefont{A.}~\bibnamefont{Ambrosetti}},
  \bibinfo{author}{\bibfnamefont{R.}~\bibnamefont{Car}}, \bibnamefont{and}
  \bibinfo{author}{\bibfnamefont{A.}~\bibnamefont{Tkatchenko}},
  \bibinfo{journal}{Phys. Rev. Lett.} \textbf{\bibinfo{volume}{114}},
  \bibinfo{pages}{176802} (\bibinfo{year}{2015}).

\bibitem[{\citenamefont{Tkatchenko et~al.}(2012)\citenamefont{Tkatchenko,
  DiStasio~Jr., Car, and Scheffler}}]{tkatchenko_accurate_2012}
\bibinfo{author}{\bibfnamefont{A.}~\bibnamefont{Tkatchenko}},
  \bibinfo{author}{\bibfnamefont{R.~A.} \bibnamefont{DiStasio~Jr.}},
  \bibinfo{author}{\bibfnamefont{R.}~\bibnamefont{Car}}, \bibnamefont{and}
  \bibinfo{author}{\bibfnamefont{M.}~\bibnamefont{Scheffler}},
  \bibinfo{journal}{Phys. Rev. Lett.} \textbf{\bibinfo{volume}{108}},
  \bibinfo{pages}{236402} (\bibinfo{year}{2012}).

\bibitem[{\citenamefont{DiStasio~Jr. et~al.}(2012)\citenamefont{DiStasio~Jr.,
  von Lilienfeld, and Tkatchenko}}]{distasio_jr._collective_2012}
\bibinfo{author}{\bibfnamefont{R.~A.} \bibnamefont{DiStasio~Jr.}},
  \bibinfo{author}{\bibfnamefont{O.~A.} \bibnamefont{von Lilienfeld}},
  \bibnamefont{and}
  \bibinfo{author}{\bibfnamefont{A.}~\bibnamefont{Tkatchenko}},
  \bibinfo{journal}{Proc. Natl. Acad. Sci. USA} \textbf{\bibinfo{volume}{109}},
  \bibinfo{pages}{14791} (\bibinfo{year}{2012}).

\bibitem[{\citenamefont{DiStasio~Jr.
  et~al.}(2014{\natexlab{a}})\citenamefont{DiStasio~Jr., Gobre, and
  Tkatchenko}}]{distasio_jr._many-body_2014}
\bibinfo{author}{\bibfnamefont{R.~A.} \bibnamefont{DiStasio~Jr.}},
  \bibinfo{author}{\bibfnamefont{V.~V.} \bibnamefont{Gobre}}, \bibnamefont{and}
  \bibinfo{author}{\bibfnamefont{A.}~\bibnamefont{Tkatchenko}},
  \bibinfo{journal}{J. Phys.: Condens. Matter} \textbf{\bibinfo{volume}{26}},
  \bibinfo{pages}{213202} (\bibinfo{year}{2014}{\natexlab{a}}).

\bibitem[{\citenamefont{Ambrosetti
  et~al.}(2014{\natexlab{a}})\citenamefont{Ambrosetti, Reilly, DiStasio~Jr.,
  and Tkatchenko}}]{ambrosetti_long-range_2014}
\bibinfo{author}{\bibfnamefont{A.}~\bibnamefont{Ambrosetti}},
  \bibinfo{author}{\bibfnamefont{A.~M.} \bibnamefont{Reilly}},
  \bibinfo{author}{\bibfnamefont{R.~A.} \bibnamefont{DiStasio~Jr.}},
  \bibnamefont{and}
  \bibinfo{author}{\bibfnamefont{A.}~\bibnamefont{Tkatchenko}},
  \bibinfo{journal}{J. Chem. Phys.} \textbf{\bibinfo{volume}{140}},
  \bibinfo{pages}{18A508} (\bibinfo{year}{2014}{\natexlab{a}}).

\bibitem[{\citenamefont{Blood-Forsythe
  et~al.}(2016)\citenamefont{Blood-Forsythe, Markovich, DiStasio~Jr., Car, and
  Aspuru-Guzik}}]{blood-forsythe_analytical_2016}
\bibinfo{author}{\bibfnamefont{M.~A.} \bibnamefont{Blood-Forsythe}},
  \bibinfo{author}{\bibfnamefont{T.}~\bibnamefont{Markovich}},
  \bibinfo{author}{\bibfnamefont{R.~A.} \bibnamefont{DiStasio~Jr.}},
  \bibinfo{author}{\bibfnamefont{R.}~\bibnamefont{Car}}, \bibnamefont{and}
  \bibinfo{author}{\bibfnamefont{A.}~\bibnamefont{Aspuru-Guzik}},
  \bibinfo{journal}{Chem. Sci.} \textbf{\bibinfo{volume}{7}},
  \bibinfo{pages}{1712} (\bibinfo{year}{2016}).

\bibitem[{\citenamefont{Dion et~al.}(2004)\citenamefont{Dion, Rydberg,
  Schr{\"o}der, Langreth, and Lundqvist}}]{dion_van_2004}
\bibinfo{author}{\bibfnamefont{M.}~\bibnamefont{Dion}},
  \bibinfo{author}{\bibfnamefont{H.}~\bibnamefont{Rydberg}},
  \bibinfo{author}{\bibfnamefont{E.}~\bibnamefont{Schr{\"o}der}},
  \bibinfo{author}{\bibfnamefont{D.~C.} \bibnamefont{Langreth}},
  \bibnamefont{and} \bibinfo{author}{\bibfnamefont{B.~I.}
  \bibnamefont{Lundqvist}}, \bibinfo{journal}{Phys. Rev. Lett.}
  \textbf{\bibinfo{volume}{92}}, \bibinfo{pages}{246401}
  (\bibinfo{year}{2004}).

\bibitem[{\citenamefont{Vydrov and Van~Voorhis}(2009)}]{vydrov_nonlocal_2009}
\bibinfo{author}{\bibfnamefont{O.~A.} \bibnamefont{Vydrov}} \bibnamefont{and}
  \bibinfo{author}{\bibfnamefont{T.}~\bibnamefont{Van~Voorhis}},
  \bibinfo{journal}{Phys. Rev. Lett.} \textbf{\bibinfo{volume}{103}},
  \bibinfo{pages}{063004} (\bibinfo{year}{2009}).

\bibitem[{\citenamefont{Lee et~al.}(2010)\citenamefont{Lee, Murray, Kong,
  Lundqvist, and Langreth}}]{lee_higher-accuracy_2010}
\bibinfo{author}{\bibfnamefont{K.}~\bibnamefont{Lee}},
  \bibinfo{author}{\bibfnamefont{{\'E}.~D.} \bibnamefont{Murray}},
  \bibinfo{author}{\bibfnamefont{L.}~\bibnamefont{Kong}},
  \bibinfo{author}{\bibfnamefont{B.~I.} \bibnamefont{Lundqvist}},
  \bibnamefont{and} \bibinfo{author}{\bibfnamefont{D.~C.}
  \bibnamefont{Langreth}}, \bibinfo{journal}{Phys. Rev. B}
  \textbf{\bibinfo{volume}{82}}, \bibinfo{pages}{081101}
  (\bibinfo{year}{2010}).

\bibitem[{\citenamefont{Becke}(1993)}]{becke_new_1993}
\bibinfo{author}{\bibfnamefont{A.~D.} \bibnamefont{Becke}},
  \bibinfo{journal}{J. Chem. Phys.} \textbf{\bibinfo{volume}{98}},
  \bibinfo{pages}{1372} (\bibinfo{year}{1993}).

\bibitem[{\citenamefont{Fosdick}(1962)}]{fosdick_numerical_1962}
\bibinfo{author}{\bibfnamefont{L.~D.} \bibnamefont{Fosdick}},
  \bibinfo{journal}{J. Math Phys.} \textbf{\bibinfo{volume}{3}},
  \bibinfo{pages}{1251} (\bibinfo{year}{1962}).

\bibitem[{\citenamefont{Chandler and Wolynes}(1981)}]{chandler_exploiting_1981}
\bibinfo{author}{\bibfnamefont{D.}~\bibnamefont{Chandler}} \bibnamefont{and}
  \bibinfo{author}{\bibfnamefont{P.~G.} \bibnamefont{Wolynes}},
  \bibinfo{journal}{J. Chem. Phys.} \textbf{\bibinfo{volume}{74}},
  \bibinfo{pages}{4078} (\bibinfo{year}{1981}).

\bibitem[{\citenamefont{Marx and Parrinello}(1996)}]{marx_ab_1996}
\bibinfo{author}{\bibfnamefont{D.}~\bibnamefont{Marx}} \bibnamefont{and}
  \bibinfo{author}{\bibfnamefont{M.}~\bibnamefont{Parrinello}},
  \bibinfo{journal}{J. Chem. Phys.} \textbf{\bibinfo{volume}{104}},
  \bibinfo{pages}{4077} (\bibinfo{year}{1996}).

\bibitem[{\citenamefont{Tuckerman et~al.}(1996)\citenamefont{Tuckerman, Marx,
  Klein, and Parrinello}}]{tuckerman_efficient_1996}
\bibinfo{author}{\bibfnamefont{M.~E.} \bibnamefont{Tuckerman}},
  \bibinfo{author}{\bibfnamefont{D.}~\bibnamefont{Marx}},
  \bibinfo{author}{\bibfnamefont{M.~L.} \bibnamefont{Klein}}, \bibnamefont{and}
  \bibinfo{author}{\bibfnamefont{M.}~\bibnamefont{Parrinello}},
  \bibinfo{journal}{J. Chem. Phys.} \textbf{\bibinfo{volume}{104}},
  \bibinfo{pages}{5579} (\bibinfo{year}{1996}).

\bibitem[{\citenamefont{Ceriotti et~al.}(2011)\citenamefont{Ceriotti,
  Manolopoulos, and Parrinello}}]{ceriotti_accelerating_2011}
\bibinfo{author}{\bibfnamefont{M.}~\bibnamefont{Ceriotti}},
  \bibinfo{author}{\bibfnamefont{D.~E.} \bibnamefont{Manolopoulos}},
  \bibnamefont{and}
  \bibinfo{author}{\bibfnamefont{M.}~\bibnamefont{Parrinello}},
  \bibinfo{journal}{J. Chem. Phys.} \textbf{\bibinfo{volume}{134}},
  \bibinfo{pages}{084104} (\bibinfo{year}{2011}).

\bibitem[{\citenamefont{Ceriotti et~al.}(2014)\citenamefont{Ceriotti, More, and
  Manolopoulos}}]{ceriotti_i-pi:_2014}
\bibinfo{author}{\bibfnamefont{M.}~\bibnamefont{Ceriotti}},
  \bibinfo{author}{\bibfnamefont{J.}~\bibnamefont{More}}, \bibnamefont{and}
  \bibinfo{author}{\bibfnamefont{D.~E.} \bibnamefont{Manolopoulos}},
  \bibinfo{journal}{Comput. Phys. Commun.} \textbf{\bibinfo{volume}{185}},
  \bibinfo{pages}{1019} (\bibinfo{year}{2014}).

\bibitem[{\citenamefont{Mootz and Wussow}(1981)}]{mootz_crystal_1981}
\bibinfo{author}{\bibfnamefont{D.}~\bibnamefont{Mootz}} \bibnamefont{and}
  \bibinfo{author}{\bibfnamefont{H.-G.} \bibnamefont{Wussow}},
  \bibinfo{journal}{J. Chem. Phys.} \textbf{\bibinfo{volume}{75}},
  \bibinfo{pages}{1517} (\bibinfo{year}{1981}).

\bibitem[{\citenamefont{Crawford et~al.}(2009)\citenamefont{Crawford, Kirchner,
  Bl{\"a}ser, Boese, David, Dawson, Gehrke, Ibberson, Marshall, Parsons
  et~al.}}]{crawford_isotopic_2009}
\bibinfo{author}{\bibfnamefont{S.}~\bibnamefont{Crawford}},
  \bibinfo{author}{\bibfnamefont{M.~T.} \bibnamefont{Kirchner}},
  \bibinfo{author}{\bibfnamefont{D.}~\bibnamefont{Bl{\"a}ser}},
  \bibinfo{author}{\bibfnamefont{R.}~\bibnamefont{Boese}},
  \bibinfo{author}{\bibfnamefont{W.~I.~F.} \bibnamefont{David}},
  \bibinfo{author}{\bibfnamefont{A.}~\bibnamefont{Dawson}},
  \bibinfo{author}{\bibfnamefont{A.}~\bibnamefont{Gehrke}},
  \bibinfo{author}{\bibfnamefont{R.~M.} \bibnamefont{Ibberson}},
  \bibinfo{author}{\bibfnamefont{W.~G.} \bibnamefont{Marshall}},
  \bibinfo{author}{\bibfnamefont{S.}~\bibnamefont{Parsons}},
  \bibnamefont{et~al.}, \bibinfo{journal}{Angew. Chem. Int. Ed.}
  \textbf{\bibinfo{volume}{48}}, \bibinfo{pages}{755} (\bibinfo{year}{2009}).

\bibitem[{\citenamefont{Goddard et~al.}(1997)\citenamefont{Goddard, Heinemann,
  and Kr{\"u}ger}}]{goddard_pyrrole_1997}
\bibinfo{author}{\bibfnamefont{R.}~\bibnamefont{Goddard}},
  \bibinfo{author}{\bibfnamefont{O.}~\bibnamefont{Heinemann}},
  \bibnamefont{and}
  \bibinfo{author}{\bibfnamefont{C.}~\bibnamefont{Kr{\"u}ger}},
  \bibinfo{journal}{Acta Crystallogr., Sect. C: Cryst. Struct. Commun.}
  \textbf{\bibinfo{volume}{53}}, \bibinfo{pages}{1846} (\bibinfo{year}{1997}).

\bibitem[{\citenamefont{Podsiad{\l }o et~al.}(2010)\citenamefont{Podsiad{\l }o,
  Jak{\'o}bek, and Katrusiak}}]{podsiadlo_density_2010}
\bibinfo{author}{\bibfnamefont{M.}~\bibnamefont{Podsiad{\l }o}},
  \bibinfo{author}{\bibfnamefont{K.}~\bibnamefont{Jak{\'o}bek}},
  \bibnamefont{and}
  \bibinfo{author}{\bibfnamefont{A.}~\bibnamefont{Katrusiak}},
  \bibinfo{journal}{Cryst. Eng. Comm.} \textbf{\bibinfo{volume}{12}},
  \bibinfo{pages}{2561} (\bibinfo{year}{2010}).

\bibitem[{\citenamefont{K{\"u}hn et~al.}(2002)\citenamefont{K{\"u}hn, Groarke,
  Bencze, Herdtweck, Prazeres, Santos, Calhorda, Rom{\~a}o, Gon{\c c}alves,
  Lopes et~al.}}]{kuhn_octahedral_2002}
\bibinfo{author}{\bibfnamefont{F.~E.} \bibnamefont{K{\"u}hn}},
  \bibinfo{author}{\bibfnamefont{M.}~\bibnamefont{Groarke}},
  \bibinfo{author}{\bibfnamefont{{\'E}.}~\bibnamefont{Bencze}},
  \bibinfo{author}{\bibfnamefont{E.}~\bibnamefont{Herdtweck}},
  \bibinfo{author}{\bibfnamefont{A.}~\bibnamefont{Prazeres}},
  \bibinfo{author}{\bibfnamefont{A.~M.} \bibnamefont{Santos}},
  \bibinfo{author}{\bibfnamefont{M.~J.} \bibnamefont{Calhorda}},
  \bibinfo{author}{\bibfnamefont{C.~C.} \bibnamefont{Rom{\~a}o}},
  \bibinfo{author}{\bibfnamefont{I.~S.} \bibnamefont{Gon{\c c}alves}},
  \bibinfo{author}{\bibfnamefont{A.~D.} \bibnamefont{Lopes}},
  \bibnamefont{et~al.}, \bibinfo{journal}{Chem. Eur. J.}
  \textbf{\bibinfo{volume}{8}}, \bibinfo{pages}{2370} (\bibinfo{year}{2002}).

\bibitem[{\citenamefont{Gilchrist}(2007)}]{gilchrist_heterocyclic_2007}
\bibinfo{author}{\bibnamefont{Gilchrist}}, \emph{\bibinfo{title}{Heterocyclic
  {Chemistry}}} (\bibinfo{publisher}{Pearson Education},
  \bibinfo{address}{Essex}, \bibinfo{year}{2007}).

\bibitem[{\citenamefont{Perdew et~al.}(1996{\natexlab{a}})\citenamefont{Perdew,
  Burke, and Ernzerhof}}]{perdew_generalized_1996}
\bibinfo{author}{\bibfnamefont{J.~P.} \bibnamefont{Perdew}},
  \bibinfo{author}{\bibfnamefont{K.}~\bibnamefont{Burke}}, \bibnamefont{and}
  \bibinfo{author}{\bibfnamefont{M.}~\bibnamefont{Ernzerhof}},
  \bibinfo{journal}{Phys. Rev. Lett.} \textbf{\bibinfo{volume}{77}},
  \bibinfo{pages}{3865} (\bibinfo{year}{1996}{\natexlab{a}}).

\bibitem[{\citenamefont{DiStasio~Jr.
  et~al.}(2014{\natexlab{b}})\citenamefont{DiStasio~Jr., Santra, Li, Wu, and
  Car}}]{distasio_jr._individual_2014}
\bibinfo{author}{\bibfnamefont{R.~A.} \bibnamefont{DiStasio~Jr.}},
  \bibinfo{author}{\bibfnamefont{B.}~\bibnamefont{Santra}},
  \bibinfo{author}{\bibfnamefont{Z.}~\bibnamefont{Li}},
  \bibinfo{author}{\bibfnamefont{X.}~\bibnamefont{Wu}}, \bibnamefont{and}
  \bibinfo{author}{\bibfnamefont{R.}~\bibnamefont{Car}}, \bibinfo{journal}{J.
  Chem. Phys.} \textbf{\bibinfo{volume}{141}}, \bibinfo{pages}{084502}
  (\bibinfo{year}{2014}{\natexlab{b}}).

\bibitem[{\citenamefont{Al-Saidi et~al.}(2012)\citenamefont{Al-Saidi, Voora,
  and Jordan}}]{al-saidi_assessment_2012}
\bibinfo{author}{\bibfnamefont{W.~A.} \bibnamefont{Al-Saidi}},
  \bibinfo{author}{\bibfnamefont{V.~K.} \bibnamefont{Voora}}, \bibnamefont{and}
  \bibinfo{author}{\bibfnamefont{K.~D.} \bibnamefont{Jordan}},
  \bibinfo{journal}{J. Chem. Theory Comput.} \textbf{\bibinfo{volume}{8}},
  \bibinfo{pages}{1503} (\bibinfo{year}{2012}).

\bibitem[{\citenamefont{Bu{\v c}ko et~al.}(2013)\citenamefont{Bu{\v c}ko,
  Leb{\`e}gue, Hafner, and {\'A}ngy{\'a}n}}]{bucko_tkatchenko-scheffler_2013}
\bibinfo{author}{\bibfnamefont{T.}~\bibnamefont{Bu{\v c}ko}},
  \bibinfo{author}{\bibfnamefont{S.}~\bibnamefont{Leb{\`e}gue}},
  \bibinfo{author}{\bibfnamefont{J.}~\bibnamefont{Hafner}}, \bibnamefont{and}
  \bibinfo{author}{\bibfnamefont{J.~G.} \bibnamefont{{\'A}ngy{\'a}n}},
  \bibinfo{journal}{Phys. Rev. B} \textbf{\bibinfo{volume}{87}},
  \bibinfo{pages}{064110} (\bibinfo{year}{2013}).

\bibitem[{\citenamefont{Reilly and
  Tkatchenko}(2013)}]{reilly_understanding_2013}
\bibinfo{author}{\bibfnamefont{A.~M.} \bibnamefont{Reilly}} \bibnamefont{and}
  \bibinfo{author}{\bibfnamefont{A.}~\bibnamefont{Tkatchenko}},
  \bibinfo{journal}{J. Chem. Phys.} \textbf{\bibinfo{volume}{139}},
  \bibinfo{pages}{024705} (\bibinfo{year}{2013}).

\bibitem[{\citenamefont{Car and Parrinello}(1985)}]{car_unified_1985}
\bibinfo{author}{\bibfnamefont{R.}~\bibnamefont{Car}} \bibnamefont{and}
  \bibinfo{author}{\bibfnamefont{M.}~\bibnamefont{Parrinello}},
  \bibinfo{journal}{Phys. Rev. Lett.} \textbf{\bibinfo{volume}{55}},
  \bibinfo{pages}{2471} (\bibinfo{year}{1985}).

\bibitem[{\citenamefont{Tobias et~al.}(1993)\citenamefont{Tobias, Martyna, and
  Klein}}]{tobias_molecular_1993}
\bibinfo{author}{\bibfnamefont{D.~J.} \bibnamefont{Tobias}},
  \bibinfo{author}{\bibfnamefont{G.~J.} \bibnamefont{Martyna}},
  \bibnamefont{and} \bibinfo{author}{\bibfnamefont{M.~L.} \bibnamefont{Klein}},
  \bibinfo{journal}{J. Phys. Chem.} \textbf{\bibinfo{volume}{97}},
  \bibinfo{pages}{12959} (\bibinfo{year}{1993}).

\bibitem[{\citenamefont{Parrinello and Rahman}(1980)}]{parrinello_crystal_1980}
\bibinfo{author}{\bibfnamefont{M.}~\bibnamefont{Parrinello}} \bibnamefont{and}
  \bibinfo{author}{\bibfnamefont{A.}~\bibnamefont{Rahman}},
  \bibinfo{journal}{Phys. Rev. Lett.} \textbf{\bibinfo{volume}{45}},
  \bibinfo{pages}{1196} (\bibinfo{year}{1980}).

\bibitem[{\citenamefont{Giannozzi et~al.}(2009)\citenamefont{Giannozzi, Baroni,
  Bonini, Calandra, Car, Cavazzoni, Ceresoli, Chiarotti, Cococcioni, Dabo
  et~al.}}]{giannozzi_quantum_2009}
\bibinfo{author}{\bibfnamefont{P.}~\bibnamefont{Giannozzi}},
  \bibinfo{author}{\bibfnamefont{S.}~\bibnamefont{Baroni}},
  \bibinfo{author}{\bibfnamefont{N.}~\bibnamefont{Bonini}},
  \bibinfo{author}{\bibfnamefont{M.}~\bibnamefont{Calandra}},
  \bibinfo{author}{\bibfnamefont{R.}~\bibnamefont{Car}},
  \bibinfo{author}{\bibfnamefont{C.}~\bibnamefont{Cavazzoni}},
  \bibinfo{author}{\bibfnamefont{D.}~\bibnamefont{Ceresoli}},
  \bibinfo{author}{\bibfnamefont{G.~L.} \bibnamefont{Chiarotti}},
  \bibinfo{author}{\bibfnamefont{M.}~\bibnamefont{Cococcioni}},
  \bibinfo{author}{\bibfnamefont{I.}~\bibnamefont{Dabo}}, \bibnamefont{et~al.},
  \bibinfo{journal}{J. Phys.: Condens. Matter} \textbf{\bibinfo{volume}{21}},
  \bibinfo{pages}{395502} (\bibinfo{year}{2009}).

\bibitem[{\citenamefont{Giannozzi et~al.}(2017)\citenamefont{Giannozzi,
  Andreussi, Brumme, Bunau, Nardelli, Calandra, Car, Cavazzoni, {D Ceresoli},
  Cococcioni et~al.}}]{giannozzi_advanced_2017}
\bibinfo{author}{\bibfnamefont{P.}~\bibnamefont{Giannozzi}},
  \bibinfo{author}{\bibfnamefont{O.}~\bibnamefont{Andreussi}},
  \bibinfo{author}{\bibfnamefont{T.}~\bibnamefont{Brumme}},
  \bibinfo{author}{\bibfnamefont{O.}~\bibnamefont{Bunau}},
  \bibinfo{author}{\bibfnamefont{M.~B.} \bibnamefont{Nardelli}},
  \bibinfo{author}{\bibfnamefont{M.}~\bibnamefont{Calandra}},
  \bibinfo{author}{\bibfnamefont{R.}~\bibnamefont{Car}},
  \bibinfo{author}{\bibfnamefont{C.}~\bibnamefont{Cavazzoni}},
  \bibinfo{author}{\bibnamefont{{D Ceresoli}}},
  \bibinfo{author}{\bibfnamefont{M.}~\bibnamefont{Cococcioni}},
  \bibnamefont{et~al.}, \bibinfo{journal}{J. Phys.: Condens. Matter}
  \textbf{\bibinfo{volume}{29}}, \bibinfo{pages}{465901}
  (\bibinfo{year}{2017}).

\bibitem[{\citenamefont{Bernasconi et~al.}(1995)\citenamefont{Bernasconi,
  Chiarotti, Focher, Scandolo, Tosatti, and
  Parrinello}}]{bernasconi_first-principle-constant_1995}
\bibinfo{author}{\bibfnamefont{M.}~\bibnamefont{Bernasconi}},
  \bibinfo{author}{\bibfnamefont{G.}~\bibnamefont{Chiarotti}},
  \bibinfo{author}{\bibfnamefont{P.}~\bibnamefont{Focher}},
  \bibinfo{author}{\bibfnamefont{S.}~\bibnamefont{Scandolo}},
  \bibinfo{author}{\bibfnamefont{E.}~\bibnamefont{Tosatti}}, \bibnamefont{and}
  \bibinfo{author}{\bibfnamefont{M.}~\bibnamefont{Parrinello}},
  \bibinfo{journal}{J. Phys. Chem. Solids} \textbf{\bibinfo{volume}{56}},
  \bibinfo{pages}{501} (\bibinfo{year}{1995}).

\bibitem[{sup()}]{sup-all}
\bibinfo{note}{See Supplemental Material at [URL will be inserted by publisher]
  for more details.}

\bibitem[{\citenamefont{Fortes et~al.}(2011)\citenamefont{Fortes, Suard, and
  Knight}}]{fortes_negative_2011}
\bibinfo{author}{\bibfnamefont{A.~D.} \bibnamefont{Fortes}},
  \bibinfo{author}{\bibfnamefont{E.}~\bibnamefont{Suard}}, \bibnamefont{and}
  \bibinfo{author}{\bibfnamefont{K.~S.} \bibnamefont{Knight}},
  \bibinfo{journal}{Science} \textbf{\bibinfo{volume}{331}},
  \bibinfo{pages}{742} (\bibinfo{year}{2011}).

\bibitem[{\citenamefont{Landau and Lifshitz}(1996)}]{landau_statistical_1996}
\bibinfo{author}{\bibfnamefont{L.~D.} \bibnamefont{Landau}} \bibnamefont{and}
  \bibinfo{author}{\bibfnamefont{E.~M.} \bibnamefont{Lifshitz}},
  \emph{\bibinfo{title}{Statistical {Physics}}} (\bibinfo{publisher}{Elsevier},
  \bibinfo{address}{Singapore}, \bibinfo{year}{1996}).

\bibitem[{\citenamefont{Heseltine and
  Elliot}(1962)}]{heseltine_thesis_determination_1962}
\bibinfo{author}{\bibfnamefont{J.~C.~W.} \bibnamefont{Heseltine}}
  \bibnamefont{and} \bibinfo{author}{\bibfnamefont{D.~W.}
  \bibnamefont{Elliot}}, Master's thesis, \bibinfo{school}{U.S. Naval
  Postgraduate School, Monterey, California} (\bibinfo{year}{1962}).

\bibitem[{\citenamefont{Ceriotti et~al.}(2009)\citenamefont{Ceriotti, Bussi,
  and Parrinello}}]{ceriotti_nuclear_2009}
\bibinfo{author}{\bibfnamefont{M.}~\bibnamefont{Ceriotti}},
  \bibinfo{author}{\bibfnamefont{G.}~\bibnamefont{Bussi}}, \bibnamefont{and}
  \bibinfo{author}{\bibfnamefont{M.}~\bibnamefont{Parrinello}},
  \bibinfo{journal}{Phys. Rev. Lett.} \textbf{\bibinfo{volume}{103}},
  \bibinfo{pages}{30603} (\bibinfo{year}{2009}).

\bibitem[{\citenamefont{Axilrod and Teller}(1943)}]{axilrod_interaction_1943}
\bibinfo{author}{\bibfnamefont{B.~M.} \bibnamefont{Axilrod}} \bibnamefont{and}
  \bibinfo{author}{\bibfnamefont{E.}~\bibnamefont{Teller}},
  \bibinfo{journal}{J. Chem. Phys.} \textbf{\bibinfo{volume}{11}},
  \bibinfo{pages}{299} (\bibinfo{year}{1943}).

\bibitem[{\citenamefont{Muto}(1943)}]{muto_force_1943}
\bibinfo{author}{\bibfnamefont{Y.}~\bibnamefont{Muto}}, \bibinfo{journal}{Proc.
  Phys. Math. Soc. Jpn.} \textbf{\bibinfo{volume}{17}}, \bibinfo{pages}{629}
  (\bibinfo{year}{1943}).

\bibitem[{\citenamefont{Ambrosetti
  et~al.}(2014{\natexlab{b}})\citenamefont{Ambrosetti, Alf{\`e}, DiStasio~Jr.,
  and Tkatchenko}}]{ambrosetti_hard_2014}
\bibinfo{author}{\bibfnamefont{A.}~\bibnamefont{Ambrosetti}},
  \bibinfo{author}{\bibfnamefont{D.}~\bibnamefont{Alf{\`e}}},
  \bibinfo{author}{\bibfnamefont{R.~A.} \bibnamefont{DiStasio~Jr.}},
  \bibnamefont{and}
  \bibinfo{author}{\bibfnamefont{A.}~\bibnamefont{Tkatchenko}},
  \bibinfo{journal}{J. Phys. Chem. Lett.} \textbf{\bibinfo{volume}{5}},
  \bibinfo{pages}{849} (\bibinfo{year}{2014}{\natexlab{b}}).

\bibitem[{\citenamefont{Tkatchenko et~al.}(2013)\citenamefont{Tkatchenko,
  Ambrosetti, and DiStasio~Jr.}}]{tkatchenko_interatomic_2013}
\bibinfo{author}{\bibfnamefont{A.}~\bibnamefont{Tkatchenko}},
  \bibinfo{author}{\bibfnamefont{A.}~\bibnamefont{Ambrosetti}},
  \bibnamefont{and} \bibinfo{author}{\bibfnamefont{R.~A.}
  \bibnamefont{DiStasio~Jr.}}, \bibinfo{journal}{J. Chem. Phys.}
  \textbf{\bibinfo{volume}{138}}, \bibinfo{pages}{074106}
  (\bibinfo{year}{2013}).

\bibitem[{\citenamefont{Marom et~al.}(2013)\citenamefont{Marom, DiStasio~Jr.,
  Atalla, Levchenko, Reilly, Chelikowsky, Leiserowitz, and
  Tkatchenko}}]{marom_many-body_2013}
\bibinfo{author}{\bibfnamefont{N.}~\bibnamefont{Marom}},
  \bibinfo{author}{\bibfnamefont{R.~A.} \bibnamefont{DiStasio~Jr.}},
  \bibinfo{author}{\bibfnamefont{V.}~\bibnamefont{Atalla}},
  \bibinfo{author}{\bibfnamefont{S.}~\bibnamefont{Levchenko}},
  \bibinfo{author}{\bibfnamefont{A.~M.} \bibnamefont{Reilly}},
  \bibinfo{author}{\bibfnamefont{J.~R.} \bibnamefont{Chelikowsky}},
  \bibinfo{author}{\bibfnamefont{L.}~\bibnamefont{Leiserowitz}},
  \bibnamefont{and}
  \bibinfo{author}{\bibfnamefont{A.}~\bibnamefont{Tkatchenko}},
  \bibinfo{journal}{Angew. Chem. Int. Ed.} \textbf{\bibinfo{volume}{52}},
  \bibinfo{pages}{6629} (\bibinfo{year}{2013}).

\bibitem[{\citenamefont{Reilly et~al.}(2016)\citenamefont{Reilly, Cooper,
  Adjiman, Bhattacharya, Boese, Brandenburg, Bygrave, Bylsma, Campbell, Car
  et~al.}}]{reilly_report_2016}
\bibinfo{author}{\bibfnamefont{A.~M.} \bibnamefont{Reilly}},
  \bibinfo{author}{\bibfnamefont{R.~I.} \bibnamefont{Cooper}},
  \bibinfo{author}{\bibfnamefont{C.~S.} \bibnamefont{Adjiman}},
  \bibinfo{author}{\bibfnamefont{S.}~\bibnamefont{Bhattacharya}},
  \bibinfo{author}{\bibfnamefont{A.~D.} \bibnamefont{Boese}},
  \bibinfo{author}{\bibfnamefont{J.~G.} \bibnamefont{Brandenburg}},
  \bibinfo{author}{\bibfnamefont{P.~J.} \bibnamefont{Bygrave}},
  \bibinfo{author}{\bibfnamefont{R.}~\bibnamefont{Bylsma}},
  \bibinfo{author}{\bibfnamefont{J.~E.} \bibnamefont{Campbell}},
  \bibinfo{author}{\bibfnamefont{R.}~\bibnamefont{Car}}, \bibnamefont{et~al.},
  \bibinfo{journal}{Acta Crystallogr., Sect. B: Struct. Sci.}
  \textbf{\bibinfo{volume}{72}}, \bibinfo{pages}{439} (\bibinfo{year}{2016}).

\bibitem[{\citenamefont{Ambrosetti et~al.}(2016)\citenamefont{Ambrosetti,
  Ferri, DiStasio~Jr., and Tkatchenko}}]{ambrosetti_wavelike_2016}
\bibinfo{author}{\bibfnamefont{A.}~\bibnamefont{Ambrosetti}},
  \bibinfo{author}{\bibfnamefont{N.}~\bibnamefont{Ferri}},
  \bibinfo{author}{\bibfnamefont{R.~A.} \bibnamefont{DiStasio~Jr.}},
  \bibnamefont{and}
  \bibinfo{author}{\bibfnamefont{A.}~\bibnamefont{Tkatchenko}},
  \bibinfo{journal}{Science} \textbf{\bibinfo{volume}{351}},
  \bibinfo{pages}{1171} (\bibinfo{year}{2016}).

\bibitem[{\citenamefont{Ambrosetti et~al.}(2017)\citenamefont{Ambrosetti,
  Silvestrelli, and Tkatchenko}}]{ambrosetti_physical_2017}
\bibinfo{author}{\bibfnamefont{A.}~\bibnamefont{Ambrosetti}},
  \bibinfo{author}{\bibfnamefont{P.~L.} \bibnamefont{Silvestrelli}},
  \bibnamefont{and}
  \bibinfo{author}{\bibfnamefont{A.}~\bibnamefont{Tkatchenko}},
  \bibinfo{journal}{Phys. Rev. B} \textbf{\bibinfo{volume}{95}},
  \bibinfo{pages}{235417} (\bibinfo{year}{2017}).

\bibitem[{\citenamefont{Perdew et~al.}(1996{\natexlab{b}})\citenamefont{Perdew,
  Ernzerhof, and Burke}}]{perdew_rationale_1996}
\bibinfo{author}{\bibfnamefont{J.~P.} \bibnamefont{Perdew}},
  \bibinfo{author}{\bibfnamefont{M.}~\bibnamefont{Ernzerhof}},
  \bibnamefont{and} \bibinfo{author}{\bibfnamefont{K.}~\bibnamefont{Burke}},
  \bibinfo{journal}{J. Chem. Phys.} \textbf{\bibinfo{volume}{105}},
  \bibinfo{pages}{9982} (\bibinfo{year}{1996}{\natexlab{b}}).

\bibitem[{\citenamefont{Wu et~al.}(2009)\citenamefont{Wu, Selloni, and
  Car}}]{wu_order-n_2009}
\bibinfo{author}{\bibfnamefont{X.}~\bibnamefont{Wu}},
  \bibinfo{author}{\bibfnamefont{A.}~\bibnamefont{Selloni}}, \bibnamefont{and}
  \bibinfo{author}{\bibfnamefont{R.}~\bibnamefont{Car}},
  \bibinfo{journal}{Phys. Rev. B} \textbf{\bibinfo{volume}{79}},
  \bibinfo{pages}{085102} (\bibinfo{year}{2009}).

\bibitem[{\citenamefont{Ko et~al.}()\citenamefont{Ko, Jia, Santra, Wu, Car, and
  DiStasio~Jr.}}]{exx-draft}
\bibinfo{author}{\bibfnamefont{H.-Y.} \bibnamefont{Ko}},
  \bibinfo{author}{\bibfnamefont{J.}~\bibnamefont{Jia}},
  \bibinfo{author}{\bibfnamefont{B.}~\bibnamefont{Santra}},
  \bibinfo{author}{\bibfnamefont{X.}~\bibnamefont{Wu}},
  \bibinfo{author}{\bibfnamefont{R.}~\bibnamefont{Car}}, \bibnamefont{and}
  \bibinfo{author}{\bibfnamefont{R.~A.} \bibnamefont{DiStasio~Jr.}},
  \emph{\bibinfo{title}{Enabling large-scale hybrid density functional theory
  calculations in the condensed phase}}, \bibinfo{note}{(in preparation)}.

\bibitem[{\citenamefont{Hooton}(1955)}]{hooton_li._1955}
\bibinfo{author}{\bibfnamefont{D.~J.} \bibnamefont{Hooton}},
  \bibinfo{journal}{Phil. Mag. Ser. 7} \textbf{\bibinfo{volume}{46}},
  \bibinfo{pages}{422} (\bibinfo{year}{1955}).

\bibitem[{\citenamefont{Errea et~al.}(2013)\citenamefont{Errea, Calandra, and
  Mauri}}]{errea_first-principles_2013}
\bibinfo{author}{\bibfnamefont{I.}~\bibnamefont{Errea}},
  \bibinfo{author}{\bibfnamefont{M.}~\bibnamefont{Calandra}}, \bibnamefont{and}
  \bibinfo{author}{\bibfnamefont{F.}~\bibnamefont{Mauri}},
  \bibinfo{journal}{Phys. Rev. Lett.} \textbf{\bibinfo{volume}{111}},
  \bibinfo{pages}{177002} (\bibinfo{year}{2013}).

\bibitem[{\citenamefont{Errea et~al.}(2014)\citenamefont{Errea, Calandra, and
  Mauri}}]{errea_anharmonic_2014}
\bibinfo{author}{\bibfnamefont{I.}~\bibnamefont{Errea}},
  \bibinfo{author}{\bibfnamefont{M.}~\bibnamefont{Calandra}}, \bibnamefont{and}
  \bibinfo{author}{\bibfnamefont{F.}~\bibnamefont{Mauri}},
  \bibinfo{journal}{Phys. Rev. B} \textbf{\bibinfo{volume}{89}},
  \bibinfo{pages}{064302} (\bibinfo{year}{2014}).

\end{thebibliography}
\end{document}